\documentclass[fleqn,usenatbib]{mnras}

\usepackage{newtxtext,newtxmath}
\usepackage[normalem]{ulem}

\usepackage[T1]{fontenc}

\DeclareRobustCommand{\VAN}[3]{#2}
\let\VANthebibliography\thebibliography
\def\thebibliography{\DeclareRobustCommand{\VAN}[3]{##3}\VANthebibliography}

\usepackage{graphicx}   
\usepackage{amsmath}    
\usepackage[normalem]{ulem} 
\usepackage{comment}    
\usepackage{soul}       
\usepackage{subcaption}

\newcommand{\Msunh}{$h^{-1}\,\mbox{M}_\odot$\,}

\newcommand{\Mpch}{$h^{-1}\,\mbox{Mpc}$\,}

\newcommand{\ie}{\textit{i.e. }}
\newcommand{\eg}{\textit{e.g. }}

\title[Local Universe model]{A Local Universe model for constrained simulations}

\author[S. Pfeifer et al.]{Simon Pfeifer$^{1}$\thanks{E-mail: spfeifer@aip.de}, Aurélien Valade$^{1,2}$,
Stefan Gottlöber$^1$, Yehuda Hoffman$^3$, Noam I. Libeskind$^{1,2}$\thanks{E-mail: nlibeskind@aip.de},\newauthor Wojciech A. Hellwing$^4$\\
$^{1}$Leibniz-Institut für Astrophysik Potsdam, An der Sternwarte 16, D-14482 Potsdam, Germany\\
$^{2}$Univ Lyon, Univ Claude Bernard Lyon 1, CNRS, IP2I Lyon / IN2P3, IMR 5822, F-69622, France\\
$^{3}$Racah Institute of Physics, Hebrew University, Jerusalem 91904, Israel\\
$^{4}$Center for Theoretical Physics, Polish Academy of Sciences, Al. Lotników 32/46, 02-668 Warsaw, Poland\\}
\date{Accepted XXX. Received YYY; in original form ZZZ}

\pubyear{2022}

\begin{document}
\label{firstpage}
\pagerange{\pageref{firstpage}--\pageref{lastpage}}
\maketitle
\setstcolor{red}

\begin{abstract}
The aim of cosmological simulations is to reproduce the properties of the observed Universe, serving as tools to test structure and galaxy formation models. Constrained simulations of our local cosmological region up to a few hundred \Mpch, the local Universe, are designed to reproduce the actual cosmic web of structures as observed. A question that often arises is how to judge the quality of constrained simulations against the observations of the Local Universe. Here we introduce the Local Universe model (LUM), a new methodology, whereby many constrained simulations can be judged and the ``best'' initial conditions can be identified. By characterising the Local Universe as a set of rich clusters, the model identifies haloes that serve as simulated counterparts to the observed clusters. Their merit is determined against a null hypothesis, the probability that such a counterpart could be identified in a random, unconstrained simulation. This model is applied to 100 constrained simulations using the Cosmicflows-3 data. Cluster counterparts are found for all constrained simulations, their distribution of separation from the true observed cluster position and their mass distribution are investigated. Lastly, the ``best'' constrained simulation is selected using the LUM and discussed in more detail.
\end{abstract}

\begin{keywords}
keyword1 -- keyword2 -- keyword3
\end{keywords}



\section{Introduction}
Cosmological simulations play a major role in studying the formation and evolution of the large scale structure (LSS) of the Universe. The majority of such simulations are presently conducted within the standard model of cosmology, the $\Lambda$CDM model. The aim of such simulations is to reproduce the properties of the observed Universe and the quality of the fidelity of how well the simulations recover the Universe serves as laboratory for testing structure and galaxy formation models. Cosmological simulations are expected to recover the statistical measures of the LSS such as the power spectrum, high order correlation functions and mass functions of the galaxy and underlying dark matter distributions.

Constrained simulations of our local cosmological neighborhood, referred to here as the Local Universe, are designed to reproduce the actual structure of a particular piece of the Universe, such as the cosmic web of clusters, filaments, sheets and voids. Unlike standard cosmological simulations whose initial conditions constitute of random realizations of the initial conditions \citep[\textit{e.g.}][]{angulo2012,alimi2012,Vogelsberger2014,schaye2015,mccarthy2017,pillepich2018,mtng2022}, constrained simulations are constrained by observational data pertaining to the particular patch of the Local Universe. 

Different approached essentially differ in both method and data that are used to generate the initial conditions. The \cite{hoffman1991} method was the first to be able to generate constrained initial conditions and is used in this work. Approaches built on top of this method use peculiar velocity survey data \citep{tully2008,tully2013,tully2016} which trace the gravitational potential and therefore are sensitive to the entire matter distribution. However, these peculiar velocity data are typically plagued by large uncertainties and sparse sampling as they rely on distance estimators. A relatively recent development applies Bayesian forward modelling \citep[\textit{e.g.}][]{kitaura2012,wang2014,wang2016,sawala2022,McAlpine2022} to the field of constrained simulations (see \cite{jasche2019}). These employ galaxy redshift surverys \citep[\textit{e.g.}][]{skrutskie2006,aihara2006,lavaux2011,huchra2012} which typically cover larger areas with much denser sampling. However, their downsides are the uncertainty in distance due to line-of-sight velocities (fingers of god) and more critically, that galaxies are biased tracers of the matter distribution which is challenging to account and correct for.

The continues development of constrained simulation efforts have led to their application to a wide variety of works, such as studies of the reionisation \citep{dixon2018,ocvirk2020}, clusters and their formation history \citep{sorce2016, olchanski2018,sorce2020}, galaxy distributions \citep{mathis2002,yepes2009,yepes2014,dolag2023}, the local group \citep{Libeskind2010,forero-romero2011,Libeskind2020}, magnetic fields \citep{dolag2005} and effects of modified gravity \citep{naidoo2022}.

Yet constrained simulations necessarily also incorporate an element of randomness. Random modes are required to simulate the region which are unconstrained; namely regions of space that have little data and small, non-linear scales which cannot be constrained with linear methods. Therefore both large and small scales are affected by the introduction of random modes. In this respect, many constrained simulations may be generated by changing the seed of the random fluctuation. Given that one can thus generate an infinite number of these constrained simulations, all with 
equally probable realization constrained by the observed data and the prior model, it is unclear exactly how to differentiate between multiple constrained simulations. A question that often arises is how to judge the quality of a given simulation, \ie how well does it reproduces the actual Universe \citep[e.g. see][for a similar arguement regarding the Local Group]{Carlesi2016}. Such a question arises in the context of the comparison of simulations constrained by galaxy redshift surveys and by galaxy peculiar velocities. The main purpose of this paper is therefore to introduce a new methodology whereby many constrained simulations can be judged and the ``best'' initial conditions can be identified, with an eye for future hydrodynamic simulation.

The core principle of the methodology presented in this work attempts to judge constrained simulations via null-hypotheses testing. The null-hypotheses in this case come from random simulations. By calculating a baseline from random simulations, which are completely unconstrained, one can effectively assign a statistical significance to a constrained simulation. In practice, the cluster distribution in the Local Universe is matched by identifying simulated cluster counterpart. The statistical significance of these cluster counterparts are then used to judge the constrained simulations.

This work also presents the first ever constrained simulations of the CF3 Universe -- the largest constrained simulations ever run with this method, the focus of which will be in an upcoming paper (Pfeifer et al. (in prep)).

The paper is organised as follows: Section~\ref{sec:data} presents the observation peculiar velocity data used to generate the constrained simulations and Section~\ref{sec:methods} describes the methods employed to generate the constrained initial conditions and run the constrained simulations. Section~\ref{sec:lumsection} describes how the Local Universe model (LUM) is constructed and Section~\ref{sec:results} presents the results of applying the LUM to the constrained simulations. Finally, the conclusions are presented in Section~\ref{sec:conclusion}.

\section{Cosmicflows-3 data}
\label{sec:data}
The Cosmicflows-3 catalogue consists of 17,669 individual galaxies with measured redshift, angular position and distance moduli \citep{tully2016}. Galaxies in groups and clusters, whose velocity, and thus redshift, is due largely to non linear motions are grouped by taking the arithmetic mean of their redshift and angular positions. Their errors are also reduced by taking the errors on the distance moduli of the group members and dividing by the square root of the number of members. The grouped data partially suppresses the virial motions and removes the main contribution of non-linearities in the velocity field. This grouping reduces the number of entries to 11,501 data points and this grouped Cosmicflows-3 catalogue is used henceforth, referred to as CF3. More than 95\% of the data lies within a redshift distance of $\lesssim16\,000$ km/s $\approx 160$ \Mpch.

\section{Methods}
\label{sec:methods}

\subsection{Bias Gaussianization correction}

The CF3 catalogue suffers from a variety of biases, one of the most important bias is referred to as the log-normal bias due to the the logarithmic relationship between the distance modulus and the luminosity distance. In CF3, distance moduli have normally distributed errors which become log-normal errors when converted to luminosity distance. This means that distance estimates are more likely to be overestimated even if a galaxy were to be observed many times, as the mean of these observations would not coincide with the true value. This bias also affects the derived peculiar velocities, resulting in underestimations, as these depend on the cosmological redshift and therefore the distance.

The bias gaussianization correction (BGc) method proposed by \cite{hoffman2021} aims at correcting for the log-normal bias by transforming the log-normal distribution, for distance and peculiar velocity, into a normal distribution. The main idea behind the correction is motivated by the fact that in the $\Lambda$CDM model, the radial components of peculiar velocities of galaxies in a cosmological shell are described by a Gaussian distribution with a well defined standard deviation. The BGc method transforms the measured log-normal distribution into a normal one using the fact that the median of such a transformation is conserved (while the mean is not). Lastly, the width of the normal distributions for the distance moduli and peculiar velocities are set to 0.19 and 275 km/s, relatively, consistent for a $\Lambda$CDM cosmology. For a more detailed description of the BGc method, we refer the reader to \cite{hoffman2021}.

\subsection{Constrained initial conditions}
To generate cosmological simulations that reproduce large scale structures constrained by observations, initial conditions (ICs) need to be generated which already contain this information. We follow the method of \cite{doumler2013c} to create these constrained ICs.

The process starts by reconstructing the linear density and velocity field from observations using the Wiener filter (WF). The WF has been successfully used to reconstruct the large scale, linear density and velocity field of the nearby universe from sparsely distributed observations \citep[\eg][]{hoffman2015} and the details of the method have been extensively described (see \eg \cite{hoffman1991,zaroubi1995,zaroubi1999}). In general, the WF is a Bayesian estimator that estimates the mean field given a set of uncertain data and a prior model, in this case peculiar velocity data with errors and the $\Lambda$CDM model, respectively. 


Constrained ICs generated directly from the observational data have been found to produce structures that are systematically offset from their known positions at redshift zero due to the bulk motions that shift structures away from their initial density peaks. To correct for this, the reverse Zeldovich approximation (RZA) shifts the constraints, the peculiar velocity measurement, to a position at an initial redshift using the reconstructed velocity field. Although the Zeldovich approximation breaks down at shell crossing, \ie within the non-linear regime, it has been shown to improve the resulting constrained simulations significantly \citep{doumler2013a}.

The RZA shifted constraints are used to generate a constrained realisation (CR). The aim of a constrained realisation (CR) is to generate a full Gausssian random field that contains within it a constrained density (or velocity) distribution of the early observed Local Universe. The WF does not fulfill this criterion as it returns the null field in unconstrained regions. These unconstrained regions are therefore supplemented with a random Gaussian random field via the \cite{hoffman1991} method. Finally, the CR is rescaled to an initial redshift to form the constrained ICs.

\subsection{Constrained simulations}

The constrained cosmological simulations were ran using \texttt{AREPO} \citep{arepo2019} with 512$^3$ dark matter particles and in a periodic  volume of 500 \Mpch along a side. The assumed cosmology is a flat, $\Lambda$CDM model using $\Omega_m=0.301$, $\Omega_{\Lambda}=0.709$, $n_s=0.961$, $\sigma_{8}=0.8293$ and $H_0=67.77$, consistent with \cite{planck2013}. This equates to a dark matter particle mass of $m_{\rm{DM}}=7.99\times10^{10}$ \Msunh. Constrained ICs were generated at $z=99$. The power spectra used to generate the ICs were generated using \texttt{CAMB} \citep{camb2011}. Haloes are identified with \texttt{SUBFIND} as part of \texttt{AREPO}.

A total of 100 constrained simulations were completed. The only difference between them are the seeds used in the random field of the CR used to generate the constrained ICs, with their unique seeds contained within the range [100,199]. In addition, a matching set of 100 random, unconstrained simulation with the same seeds were generated.

\begin{figure*}
    \centering
    \includegraphics[width=\textwidth]{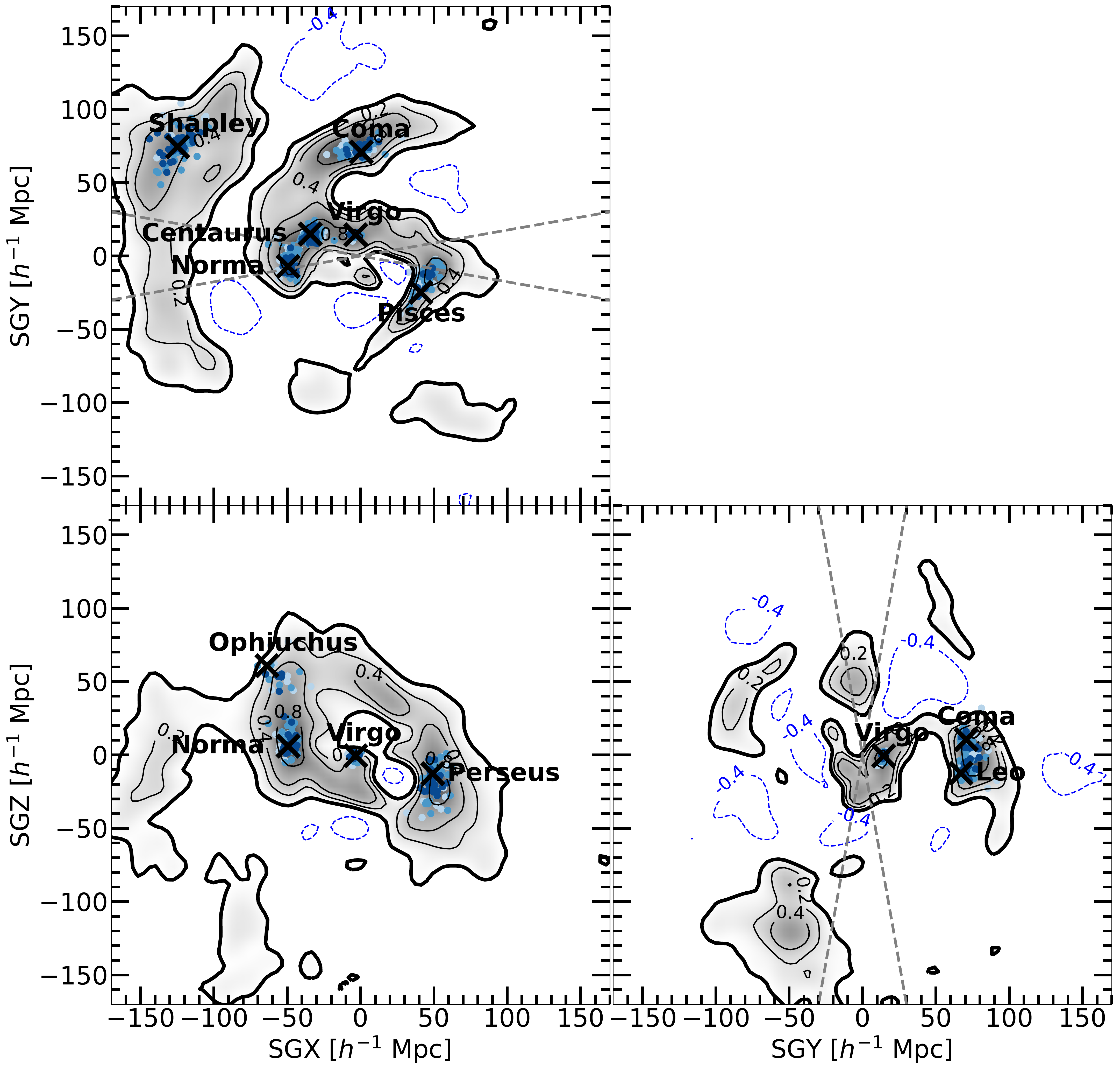}
    \caption{The contours show the quasi-linear density field, the mean of the logarithm of $\Delta=\frac{\rho}{\rho_{\rm m}}$ calculated from all 100 constrained simulations. The contours show underdensities (dashed blue lines), mean density (think black lines) and overdensity (thin black lines). The slices are centred on the supergalactic origin and 15.6 \Mpch thick. The observed cluster positions (black crosses) from Table~\ref{tab:clusterdata} are also shown. Locations of all simulated cluster counterparts (blue dots) from the constrained simulation that meet the minimum of $p<0.05$ are shown with shades indicating which $p$-value threshold they satisfy (darker is lower, Fig.~\ref{fig:clusterdetectionrate}). The simulated counterparts are projected onto the density slice irrespective of their position. The approximate location of the Zone of Avoidance, of 10 degrees, is indicated in the SGX-SGY and SGY-SGZ planes (dashed grey line).}
    \label{fig:qlcicclusters}
\end{figure*}

Fig.~\ref{fig:qlcicclusters} shows the quasi-linear (QL) density field, which is the geometric mean value taken over an ensemble  of the 100 constrained simulations \citep[following ][]{hoffman2018}. The figure shows the logarithm of $\Delta=\frac{\rho}{\rho_{\rm m}}$ smoothed on 5 \Mpch. The density slices are centred on the supergalactic origin and 15.6 \Mpch thick, going from -7.8 to +7.8 \Mpch. 

The validity of the simulations was checked at  $z=0$ by examining the total matter power spectrum and the halo mass function. The constrained and random simulations were perfectly consistent with each other as well as with the linear matter power spectrum on large scales and the \cite{tinker2008} fitted halo mass function.

\section{Local Universe model}
\label{sec:lumsection}

\subsection{Cluster selection}
\label{sec:clusterselection}
In order to judge the quality of the constrained simulations, we need a metric against which to define the quality of a given constrained simulation. In other words, it would be very desirable to have a criterion with which one can measure the likeness of a constrained simulation relative to the observed Local Universe.  We have chosen to use the distribution of massive clusters for this purpose for three main reasons; firstly, since the methods for generating constrained initial conditions rely on peculiar velocities, massive clusters with their many member galaxies produce good peculiar velocity estimates in the CF3 catalogue and therefore produce good constraints. The clusters, especially within $\approx200$ \Mpch, are observationally robust and least bias. Secondly, massive clusters are undergoing active formation today, and are therefore more closely linked to the linear regime than other collapsed structures. The overdensities they inhabit cause large scale linear flows which can be reconstructed within the linear limit of the constrained initial conditions methods. Last but not least, the distribution of clusters is among the most basic features that characterizes the specific large scale structure we inhabit. 

The cluster selection is based on the clusters in CF3 with more than $\gtrsim 50$ group members with measured velocities. This is a practical choice which reduces the error on their measured velocity. It is however also a realistic choice as all of the richest clusters in the Local Universe fulfil this criterion. The angular position in supergalactic coordinates (SGL, SGB) is very well measured but the distance to these objects can have large uncertainties depending on the method used for the distance estimation. We therefore opt to use the redshift distance since the uncertainty on the redshift measurements are negligible. Since the clusters are selected based on their large number of member galaxies, the grouping of CF3 removes the majority of the redshift space distortion, due to large small-scale velocities, by averaging the velocity measurements of the cluster members. Equation~\ref{eq:redshiftdistance} shows the equation to calculate the redshift distance,

\begin{equation}
    d = \frac{1}{H_0}(cz - v_{\rm pec})
    \label{eq:redshiftdistance}
\end{equation}

where $H_0=75.0$ km s$^{-1}$ Mpc$^{-1}$ for CF3, $c$ is the speed of light, $z$ is the redshift and $v_{\rm pec}$ is the peculiar velocity in the CMB reference frame. Column `Vcgp' in the CF3 data gives $cz-v_{\rm pec}$. Although Equation~\ref{eq:redshiftdistance} is an approximation, the error is $<1\%$ for the distances of the selected clusters.
The cluster selection and their converted supergalactic coordinates are presented in Table~\ref{tab:clusterdata}. When the simulated haloes, drawn from the constrained simulations, are compared with data, their positions are also converted to redshift distances using their radial peculiar velocities relative to a virtual observer at the centre of the box. The observed cluster masses are not explicitly considered here, other than the assumption that these clusters are more massive than $10^{14}\,h^{-1}M_\odot$.

\begin{table}
    \centering
    \begin{tabular}{l r r r}
        \hline
        Cluster & SGX     & SGY     & SGZ \\
                & [\Mpch] & [\Mpch] & [\Mpch] \\
        \hline
        Virgo & -3.3004 & 14.4336 & -0.6076 \\
        Centaurus & -34.3491 & 14.9497 & -7.5807 \\
        Hydra & -24.3204 & 20.8672 & -24.6429 \\
        Perseus & 49.1633 & -10.5576 & -12.7799 \\
        Coma & 0.4447 & 70.7675 & 10.4502 \\
        Norma & -49.4307 & -7.1231 & 6.0524 \\
        Ophiuchus & -63.7704 & 7.5590 & 60.7693 \\
        PavoII & -39.3256 & -18.5555 & 8.5548 \\
        Leo & -2.3772 & 67.0677 & -12.4381 \\
        Pisces & 41.0872 & -24.7560 & 4.5935 \\
        Shapley (A3558) & -124.7189 & 74.4359 & -3.3975 \\
    \end{tabular}
    \caption{The observational cluster data used to compare the constrained simulations against. Supergalacitc Cartesian coordinates have been calculated using the SGL, SGB and redshift distance from the CF3 grouped catalogue.}
    \label{tab:clusterdata}
\end{table}

Observational constraints on the masses of the clusters are not directly included in the LUM because of the difficulty in obtaining robust constraints in the first place. We therefore leave the nuances of observational mass estimates to a future study.

Fig.~\ref{fig:qlcicclusters} shows the relevant position of the observed cluster from Table~\ref{tab:clusterdata}, projected onto the density slices (black crosses). Note that the positions of Perseus on the SGX-SGZ plane and Coma on the SGX-SGY plane are in reality outside the region of the slice and thus small projection effect are present. The chosen cluster selection clearly traces the QL density field, generally sitting within density peaks. Ophiuchus is the only cluster in Fig~\ref{fig:qlcicclusters} that is at the boundary of the overdensity region and not within its own density peak.

\subsection{Local Universe model}
\label{sec:lum}
The WF/CR method is able to constrain regions that are well sampled by the observations. In unconstrained regions with poor sampling, and on small scales within the non-linear regime, the method must introduce random, unconstrained fluctuations. Therefore, different random initial fluctuation can produce constrained simulation of varying quality, enhancing or degrading the cosmographic features of the Local Universe. To assess the quality of constrained simulations, this work presents the LUM.

As described in Section~\ref{sec:clusterselection}, we characterise the Local Universe via a selection of observed galaxy clusters. The aim of the LUM is thus to quantify the quality of a constrained simulations with respect to the observed cluster distribution. If the separation between observation and simulation are small, one can deem the constrained simulation a good analogue of the Local Universe. Therefore, for each observed cluster, one can attempt to identify the best counterpart from the simulated halo distribution and find a quantitative measure for their similarity, such as the separation between the simulated halo and observed cluster position. However, the difficulty lies in judging the significance of this separation. To establish a baseline, the probability of identifying a cluster counterpart in an unconstrained random simulation is calculated. This can then be used to judge constrained simulations.

Firstly, only simulated haloes with M$_{200\rm c}>10^{14}$ \Msunh are considered since we are interested in rich cluster counterparts \citep{sorce2018}. The observed cluster masses are therefore not explicitly built into this method but encapsulated by this mass cut.

\begin{figure}
    \centering
    \includegraphics[width=\columnwidth]{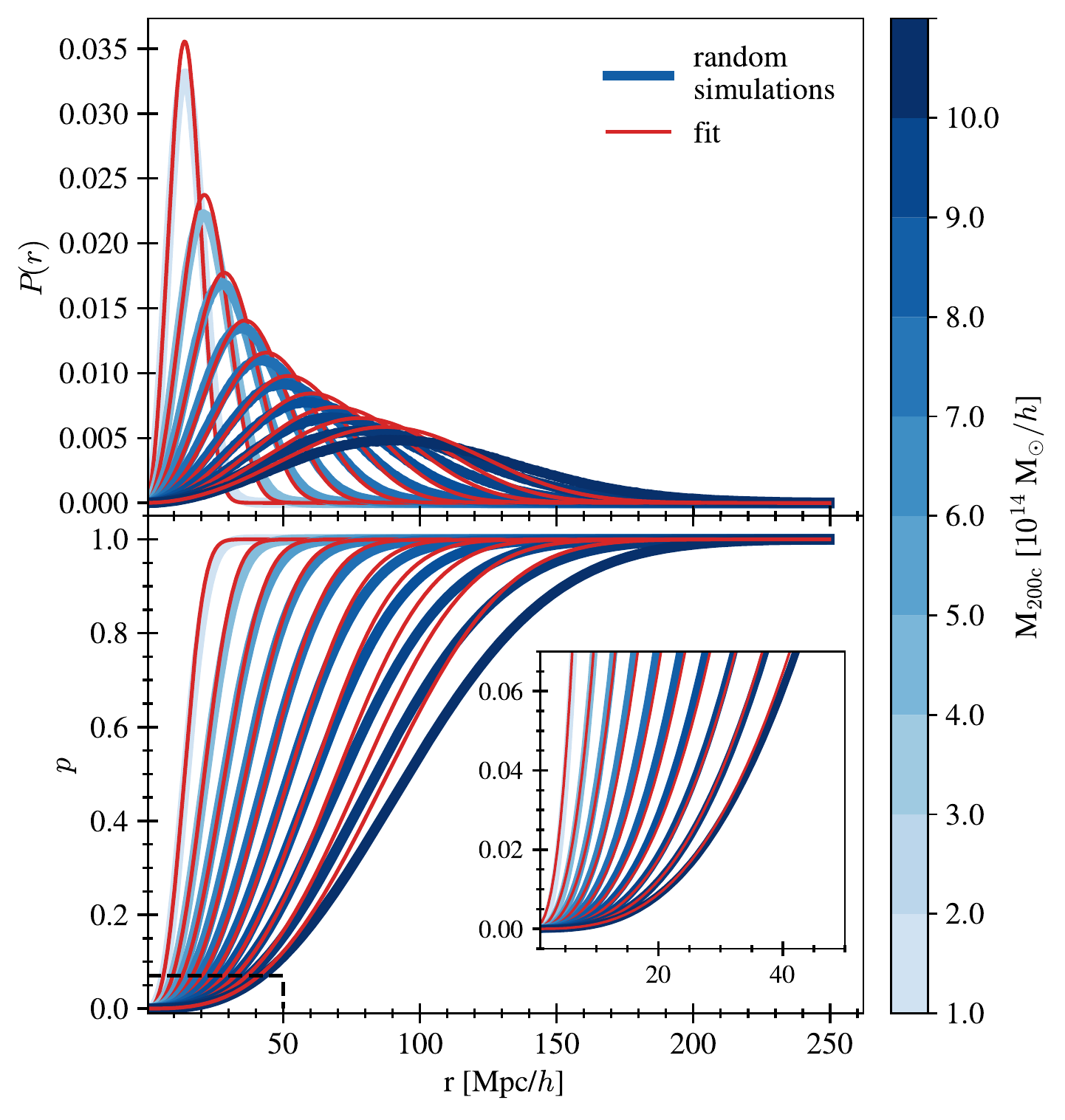}
    \caption{The PDF of the separation between the closest halo and a random point, \ie the probability of finding a halo at a given separation within a given mass cut (top). The PDFs are calculated from random, unconstrained simulations (blue) and fitted (red). The $p$-value for different mass cuts (bottom) are calculate by taking the cumulative PDF, the CDF. An insert shows a zoom-in on the distributions of $p$ for small values to show the quality of the fits in the range that is important for the LUM, represented by the black dashed area.}
    \label{fig:separationpdf}
\end{figure}

The statistic of choice for the baseline, or null-hypothesis, is the distribution of separation between observed and simulated clusters in random simulations. To measure this, a virtual observer can be placed on a location uniformly sampled across a random simulation volume and the separation between the position of, \eg Coma, and the closest halo with M$_{200\rm c}>10^{14}$ \Msunh can be measured. Although Coma occupies a special place in the Local Universe, attempting to find a halo at the position of Coma in a random simulation is no different to finding a halo at any random point in space. Similarly, one could shifting the virtual observer to a new location in the same simulation and repeat the measurement. Therefore, in practice, the separation between 10000 random points and their closest halo are measure for all 100 random simulations. Note that we assume here that the environment of the virtual observer, \ie the Milky Way, has no significant effect on the environment of clusters tens of \Mpch away. Fig.~\ref{fig:separationpdf} shows the measured probability density functions (PDF) (top) for haloes with different lower mass cuts (shaded blue lines). The centre and spread of the PDFs increases with mass because the number density of high mass haloes is lower and therefore it is less likely to find a halo at small separations. 
 
A statistical significance can now be assigned using the $p$-value. The $p$-value is generally used in null-hypothesis testing, signifying the probability that a given result, or a more extreme result, can be obtained given a null hypothesis. In this case, the random simulation provides the null-hypothesis which can be rejected with the probability of $1-p$. Of interest is the significance from zero, \ie separations of a given value and closer, and not from the mean of the PDF. Therefore the one-tailed $p$-value is used, calculated from the cumulative PDF (CDF), and shown in Fig.~\ref{fig:separationpdf} (bottom).

\begin{figure}
    \centering
    \includegraphics[width=\columnwidth]{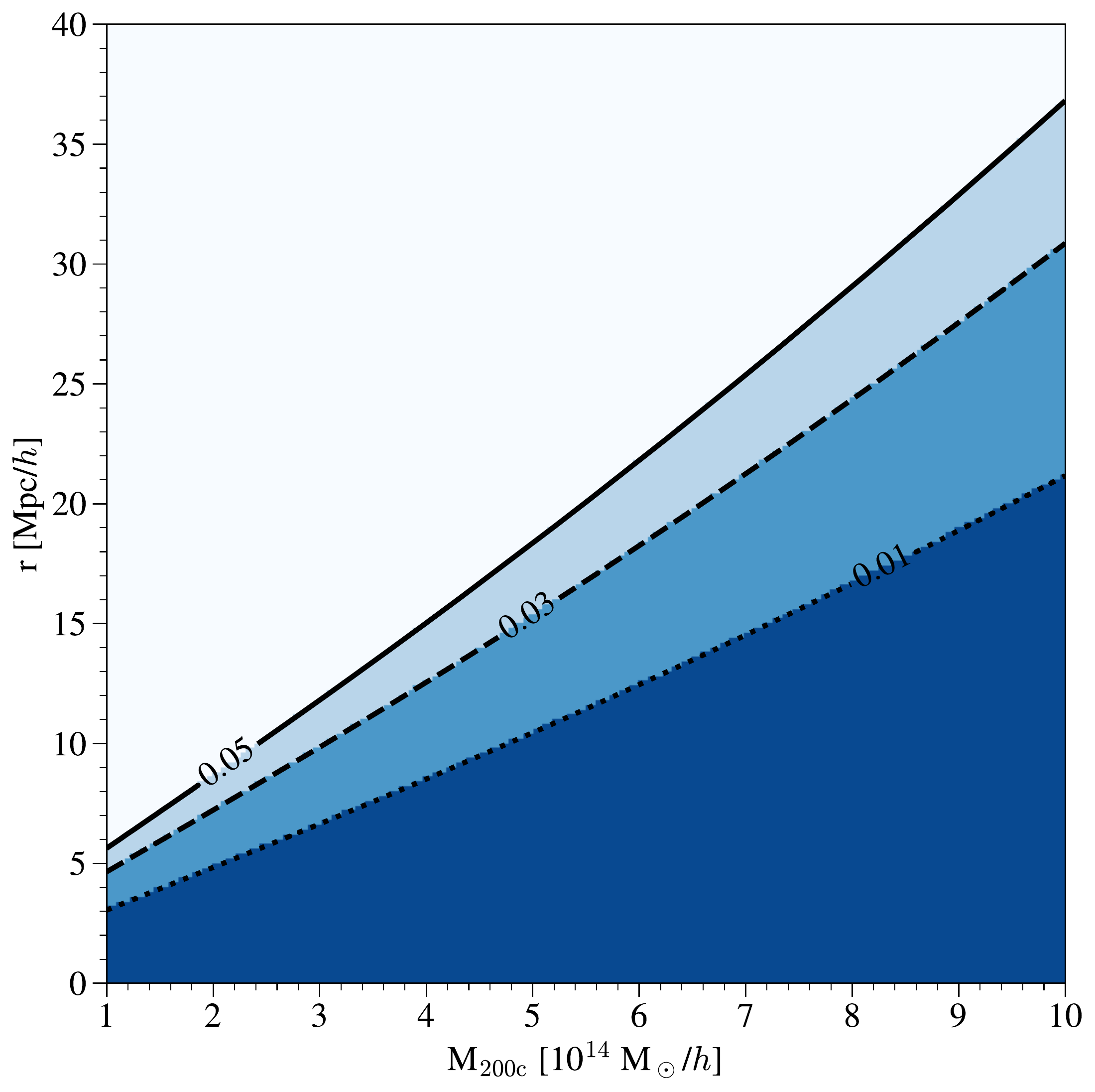}
    \caption{The $p$-values, calculated from the fits shown in Fig.~\ref{fig:separationpdf}, as a function of lower mass limit and separation. The lines denote contours of constant $p$ and darker shaded areas show lower $p$-values. The distribution above $10^{15}$ \Msunh are kept constant as not enough halos of these masses exist in the simulations to produce stable fits.}
    \label{fig:pvalsepmass}
\end{figure}

To calculate $p$ for a halo-cluster pair, the mass of the simulated halo and the separation to the observed cluster position is required. However, the CDFs in Fig.~\ref{fig:separationpdf} are calculated for a set number of mass cuts. To be able to calculate $p$ as a smooth function of separation and mass, the CDFs for each mass cut are fit using Equation~\ref{eq:analytic}, where $\alpha$ is a free parameter\footnote{The parameter $\alpha$ is actually representative of a density but for clarity and to avoid confusion with the $p$-value, we use the letter $\alpha$ instead of $\rho$ since the physical meaning is somewhat lost here. See Appendix A for more details.}, $r$ is the outer radius of a shell of fixed thickness $dr=0.5$ \Mpch. The values for $r$ and $dr$ are taken as the values used to calculate the PDFs in Fig.~\ref{fig:separationpdf}.

\begin{equation}
    P(r,\alpha) = e^{-\frac{4}{3}\pi \alpha r^3} - e^{-\frac{4}{3}\pi \alpha (r+dr)^3}
    \label{eq:analytic}
\end{equation}
The free parameter, $\alpha$, is then fit as a function of mass using a quadratic function given by Equation~\ref{eq:alpha}. \begin{multline}
    \log_{10}(\alpha) = -0.861\log_{10}(M_{\rm 200c})^2+\\
    22.616\log_{10}(M_{\rm 200c})-152.154
    \label{eq:alpha}
\end{multline}

The method uses the mass of the simulated halo to generate the corresponding PDF (normalised to sum to unity over a large range of $r$), from which the CDF and therefore $p$ can be calculated using the separation to the cluster position. The output PDFs and CDFs of these fits for the matching mass cuts are shown in Fig.~\ref{fig:separationpdf} (red lines). The up-turn of the CDFs at low separations are recovered well which is the area of interest (see inset). The fits are not accurate over the full PDF, especially the long tail at large separation. However, for the purposes of the LUM, only the region covered by the insert in Fig.~\ref{fig:separationpdf} are important. More details on the fitting of these functions is given in Appendix A. Note that it is the mass of the simulated halo and not the mass of the observed clusters that is used.

The complete pipeline effectively result in a function $p(M_{\rm 200c},r)$ and Fig.~\ref{fig:pvalsepmass} shows that relationship between the mass and separation on $p$. The lines denote contours of different $p$ with darker shades showing lower $p$-values. For a mass cut of $10^{14}$ \Msunh, a $p=0.05$ corresponds to $\approx6$ \Mpch; \ie a halo of that mass which is 6 \Mpch away from the observed position of a cluster would \textit{not} be found at that separation or closer in a random simulation with a probability of 0.95 (2$\sigma$ confidence). For the same mass cut and a separation of $\approx3$ \Mpch, this probability increase to 0.99, \ie $p=0.01$. Fig.~\ref{fig:pvalsepmass} encapsulates the essence of the LUM; it shows a metric of gauging if a simulated halo, given its mass and separation to an observed target, is significant or random. 

Note that this approach assumes that finding a cluster counterpart has no effect on the probability of finding another cluster counterpart elsewhere and thus assumes a 1-point probability distribution. The method assigns a probability to a single simulated cluster counterpart of being found in a random simulation, \ie it does not take into account the conditional probability of finding a Virgo, having found a Coma, and instead assume that each detection is independent.

\begin{figure}
    \centering
    \includegraphics[width=\columnwidth]{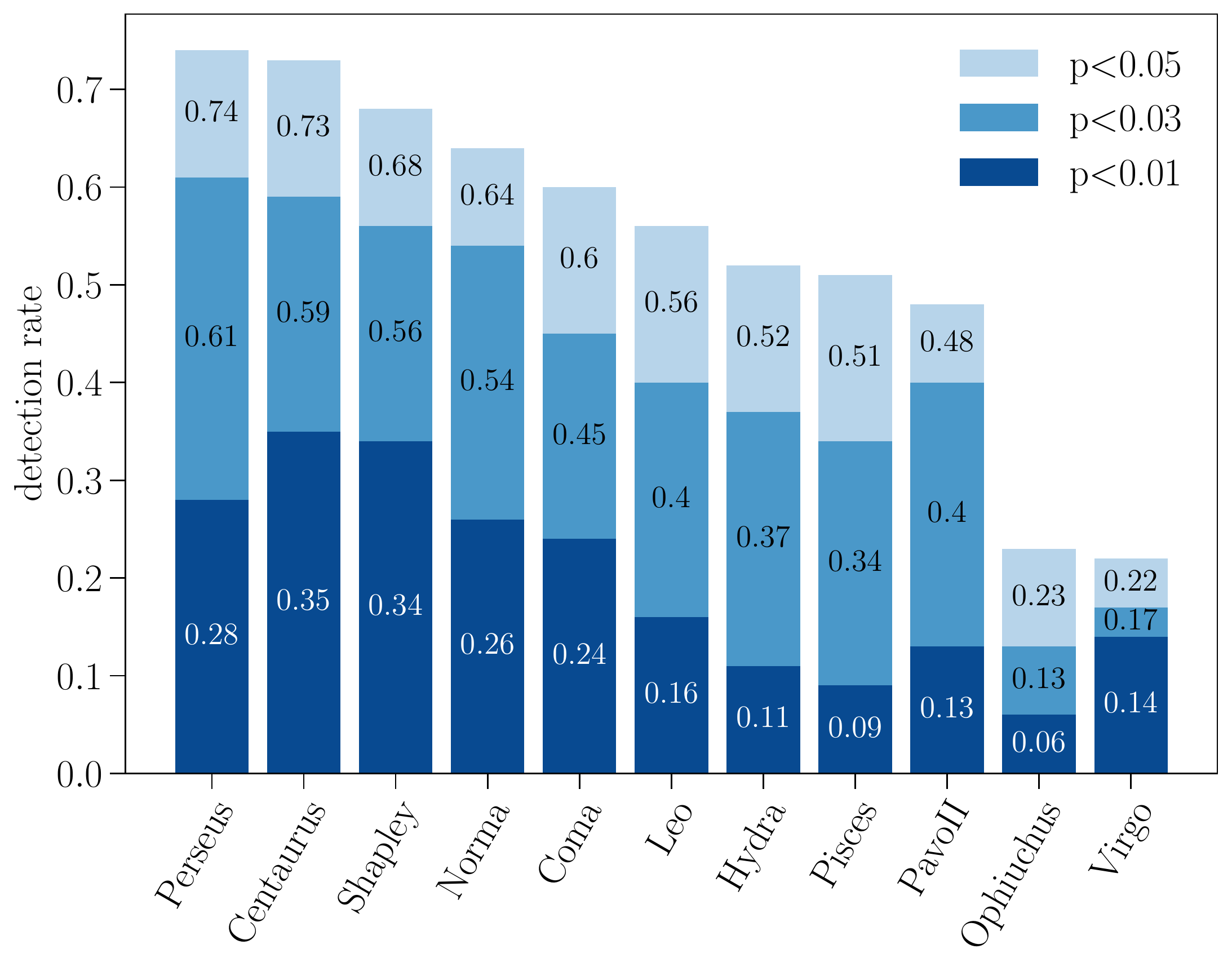}
    \caption{The fraction of detected cluster counterparts for each observed cluster from the 100 constrained simulations given different thresholds of $p$ (blue shades). Each cluster can only have one counterpart per simulation so a detection rate of 0.5 means that 50 out of 100 simulations contain a counterpart given a $p$ threshold.}
    \label{fig:clusterdetectionrate}
\end{figure}

\section{Results}
\label{sec:results}

\begin{figure*}
\centering
\begin{subfigure}{0.5\textwidth}
  \centering
  \includegraphics[width=.95\linewidth]{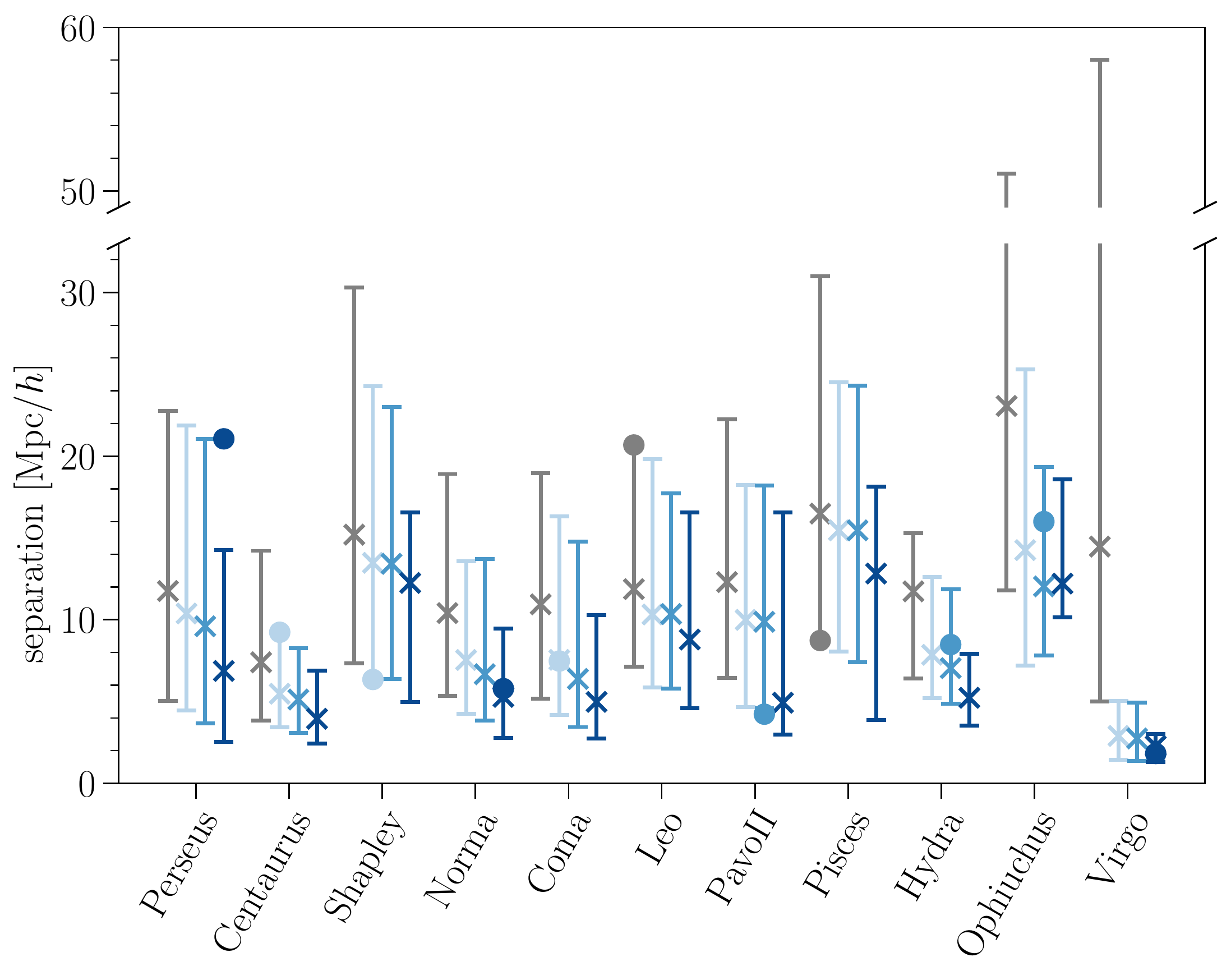}
\end{subfigure}%
\begin{subfigure}{0.5\textwidth}
  \centering
  \includegraphics[width=.95\linewidth]{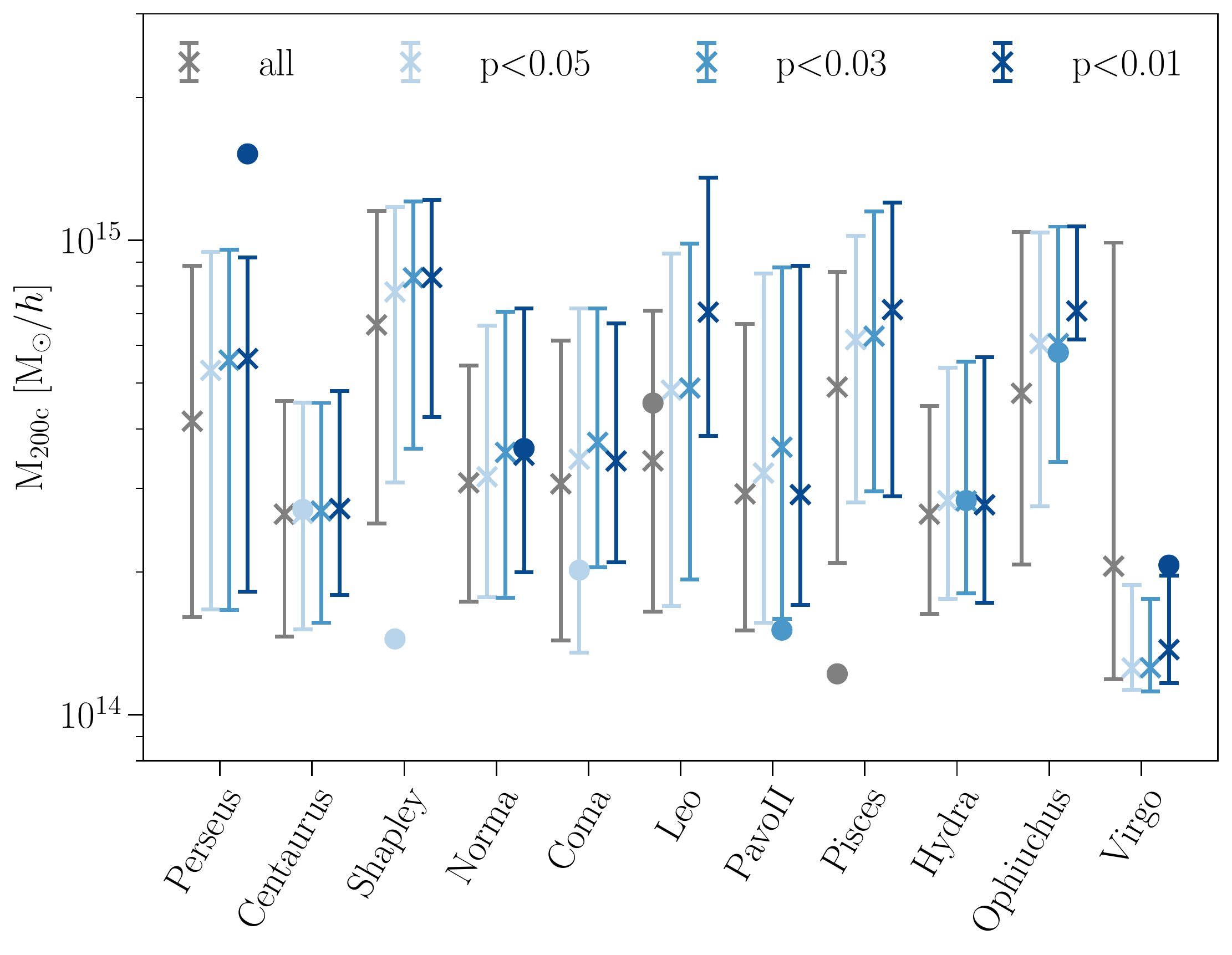}
\end{subfigure}
\caption{The median (crosses), 16th and 84th percentile (error bars) of the separation between the observed cluster position and the simulation counterpart (left) and the mass of the simulation counterparts (right). The grey line indicates the distribution for all ``best'' counterparts while the different shades indicate groups of counterparts satisfying different thresholds of $p$, where darker is lower. The circles indicate the values for the counterparts from the best constrained simulation with a seed of 159.}
\label{fig:sepmass}
\end{figure*}

\subsection{Cluster counterpart detection}
The main purpose of the LUM, presented in Section~\ref{sec:lum}, is to determining the significance of finding a halo of a given mass at a particular separation. We now leverage this method to identify the ``best'' halo counterparts to the cluster selection from CF3 for each constrained simulation.

We define ``best'' as the halo least likely to be found in a random simulation. Therefore, $p(M_{\rm 200c},r)$ can be minimised to find the ``best'' halo counterpart. For each of the $N$ observed clusters, the separation to all $M$ simulated haloes and their halo mass is used to calculate $p$ for every halo-cluster pair. This results in an $N\times M$ matrix of $p$-values. The problem here is to assign each observational cluster a unique halo counterpart, \ie each halo is only allowed to be assigned once, such that the sum of $p$ for the chosen haloes is minimised. This type of problem is referred to as an \textit{assignment problem}, which is solved efficiently with the Hungarian method \citep{hungarian}. These haloes can now be accepted or rejected as counterparts to their respective observed cluster based on their value of $p$. Fig.~\ref{fig:clusterdetectionrate} shows the fraction of identified cluster counterparts for each observed cluster from the 100 constrained simulation for three different thresholds of $p$. For the majority of clusters, the detection rates are similar. For $p<0.05$, all clusters except Ophiuchus and Virgo are found in $\gtrsim 50\%$ of the simulations. For these clusters, the trend with $p$ thresholds is consistent as well; lower thresholds reduce the detection rate as the criterion becomes more restrictive. It is interesting to see that some clusters are recovered more frequently than others. This plot effectively shows the stability of each cluster counterpart detection which is a reflection of the quality of the constraints for that particular environment. Therefore, one can expect to find a Perseus and Centaurus counterpart in most simulations.

Fig.~\ref{fig:qlcicclusters} gives a visual impression of the distribution of the counterparts with respect to the observed cluster positions. The contours show the QL density calculated from the 100 constrained simulations. All cluster counterparts for their respective cluster are shown (blue dots) with shades indicating which threshold of $p$ they satisfy, the same as for Fig.~\ref{fig:clusterdetectionrate}. The simulated counterparts are spatially distributed around their respective cluster positions, where closer counterparts tend to have lower $p$ (although the darkest points are often obscured by the cross of the cluster marker). There is a clear correlation between the distribution of counterparts and the shape of the QL overdensity that their respective observed cluster occupies. The counterpart distributions trace out the contours very well. 

Considering the first of the two outliers in Fig.~\ref{fig:clusterdetectionrate}, Ophiuchus has a low detection rate even at $p<0.05$, and only 6 counterpart are detected at $p< 0.01$. From Fig.~\ref{fig:qlcicclusters} it is clear the Ophiuchus does not sit within an overdensity in the QL field and therefore it is more likely that the best counterpart is found further away and thus have higher $p$. For Virgo, the detection rate for $p< 0.05$ is also low. Its immediate environment is within a well constrained, high overdensity region. Although the size of the region appears significantly smaller, especially apparent in the SGY-SGZ plane. The observed cluster position is directly within a QL overdensity and the spread of the simulated cluster counterparts is very tight. From Fig.~\ref{fig:qlcicclusters} is is not obvious why Virgo has such a low detection rate. This is explored in more detail in the next section. By looking at the mass distributions of the simulated counterparts it shows that the minimum mass limit of $10^{14}$ \Msunh is too high for the majority of Virgo counterparts.

\subsection{Separation and mass distribution}

The LUM uses the separations and halo masses to determine how likely an equal or better halo could be drawn from a random simulation. Fig.~\ref{fig:sepmass} shows the distributions of separations between the observed cluster and the cluster counterparts (left) and the distribution of simulated counterpart masses (right). The median (crosses), and 16th and 84th percentiles of the distribution (error bars) are shown for different thresholds of $p$, where darker shades indicate lower thresholds. The distributions for all 100 counterparts, independent of their $p$-value, are also shown (grey).

The separation distributions in Fig.~\ref{fig:sepmass} (left) change significantly with varying threshold of $p$. Generally, the median separations are $\sim$10 \Mpch for $p<0.05$ and decrease to $\sim$5 \Mpch for $p<0.01$. The spread also decreases significantly with decreasing threshold. The large changes present in the separation distributions are not found in the mass distributions. Fig.~\ref{fig:sepmass} (right) shows that the median and spread of the masses are relatively constant with varying $p$ threshold, although median masses generally increase by a small amount with decreasing threshold. 

Generally, a trade-off between mass and separation is struck due to the shape of the $p(M_{\rm 200c},r)$ distribution in Fig.~\ref{fig:pvalsepmass}; counterparts with larger separations also tend to have larger masses. Therefore, as in the case of \eg Leo, the cluster counterparts are allowed to have a large separation only if the mass increases significantly. 

Again, Virgo is an outlier with relatively small median separations and mass for all thresholds. Notably, the spread for both quantities is also significantly smaller, except for the distribution containing all counterparts (and therefore predominantly rejected counterparts) which are significantly larger. The mass distribution appears to be cut off by the lower mass limit. To test if this, the LUM was reran for a minimum mass cut of $M_{\rm 200c}>5\times10^{13}$ \Msunh. The detection rate for Virgo increase by 40\% while it increase up to $\approx10\%$ for the other clusters. The separation and mass distributions were robust to this change except that the lower percentile of the mass distribution for Virgo decreased, with a median value of $\approx10^{14}$ \Msunh. This suggests that while the overdensity of Virgo is well constrained, it struggles to form haloes of masses $>10^{14}$ \Msunh.

\subsection{Best constrained Local Universe}

\begin{figure*}
    \centering
    \includegraphics[width=\textwidth]{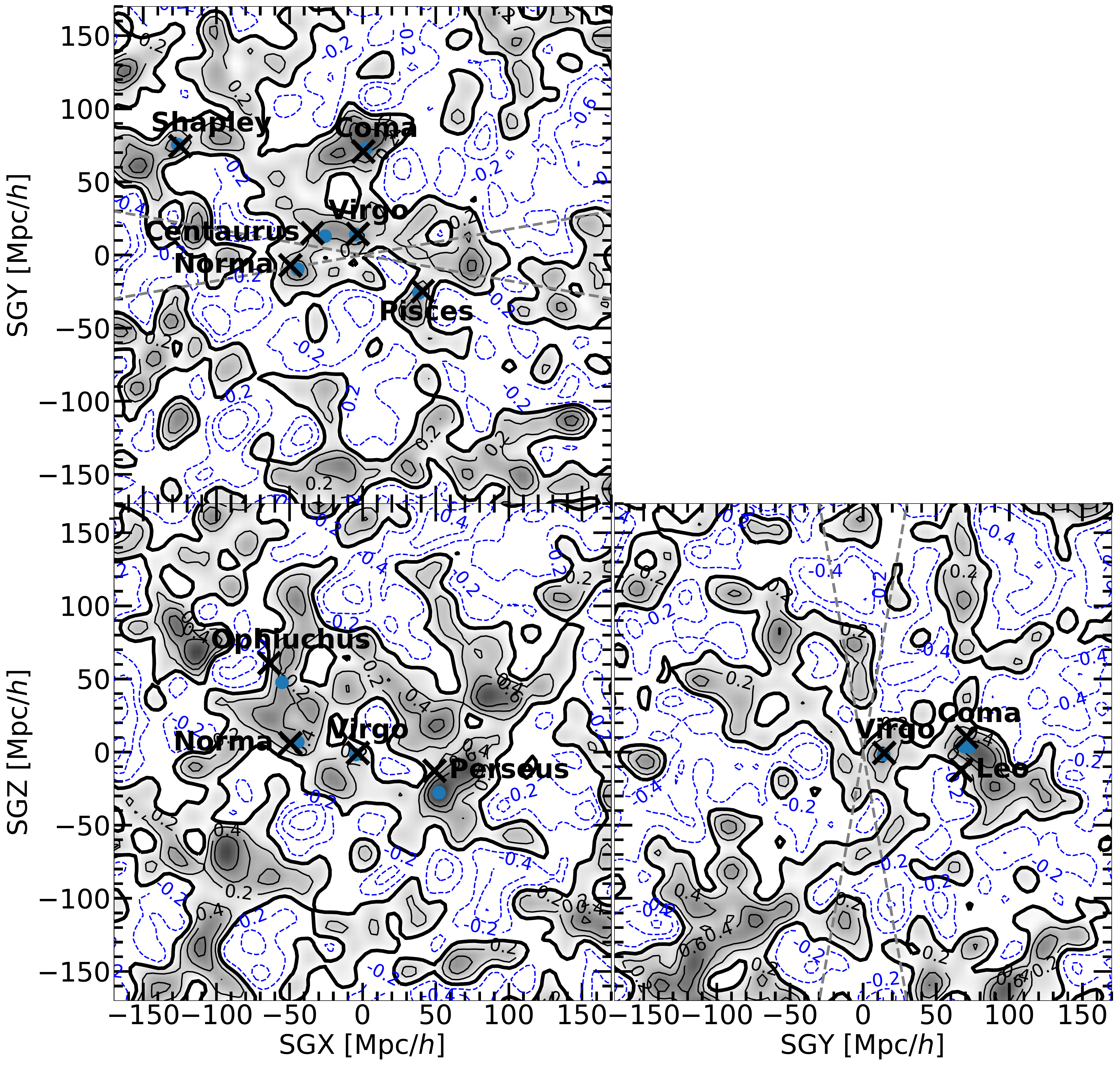}
    \caption{The contours show log$_{10}$($\Delta$) from the single constrained simulations with seed 159, the best simulation determined by the LUM. The contours show underdensities (dashed blue lines), mean density (think black lines) and overdensity (thin black lines). The density slice is centred on the supergalactic origin and 15.6 \Mpch thick. The observed cluster positions (black crosses) and the locations of all cluster counterparts (blue dots) from the 159 constrained simulation are also shown. The simulated counterparts are projected onto the density slice irrespective of their position. The approximate location of the Zone of Avoidance, of 10 degrees, is indicated in the SGX-SGY and SGY-SGZ planes (dashed grey line).}
    \label{fig:cicclusters}
\end{figure*}

The LUM can be used to find the best cluster candidate for a given simulation and a given cluster. It follows that the LUM can be used to choose the `best' candidate halo for zoom simulations of a particular cluster. Another use of the LUM is the selection of the `best' realisation for performing `full box' constrained simulations of our local volume. One way to materialize the later goal is to minimise the sum of $p$, $\Sigma p$, for all simulations, \ie the best simulation is the one with the lowest total sum of $p$ for all counterpart. Another option is to count the number of simulated cluster counterparts that are below a threshold, \eg $p<0.05$. Note that this is done under the assumption that the clusters are uncorrelated, \ie identifying a good Virgo counterpart does not affect the search for a good Coma counterpart. For the set of 100 constrained simulations, the top 5 simulations are within 0.02 of each other for $\Sigma p$ (which has the range [0.38,2.00] between the best and worse simulation, respectively) and 4 simulations have a total of 9 cluster counterparts with $p<0.05$, where 2 simulations are in both groups. The simulation with the seed of 159 (S159) has the lowest total $p$ and the most cluster counterparts with $p<0.05$, and is therefore deemed the ``best''. There is room to tailor this approach further by, \eg, placing special value on certain clusters that should always have an acceptable counterpart by enforcing that these must meet the threshold of $p<0.05$.

The locations of the cluster counterparts for S159 are shown with respect to the log$_{10}$($\Delta$) field of the constrained simulation in Fig~\ref{fig:cicclusters}. It is clear that the majority of the cluster counterparts lie very close to the observed cluster positions within each plane. The visible outliers are Perseus, Ophiuchus and Leo (which actually lies closer to Coma in the SGY-SGZ plane). For the majority of clusters in S159, the observed cluster position coincides with an overdensity peak in the constrained simulation making it likely that a massive halo is found nearby. For some, \eg Perseus, the observed cluster position sits outside or on the border of overdense regions and thus the best counterpart is further away.
It is also worth noting the striking similarities between the QL in Fig.~\ref{fig:qlcicclusters} and the S159 density features. Not only do the shapes and sizes of the  overdensities match very closely within a region of [-100, 100] \Mpch around the observer, but also the underdense void regions such as between Virgo and Perseus in the SGX-SGY plane, and in the centre of the SGX-SGZ plane.

The values of the separation of these cluster counterparts relative to the observed cluster positions are shown in Fig.~\ref{fig:sepmass} (left; circles). The same outliers are visible here as well, having separations of $>15$ \Mpch while all other cluster counterparts have separations of $<10$ \Mpch. By looking at the masses of the counterparts shown in Fig.~\ref{fig:sepmass} (right; circles), it is possible to see again the trade-off between separation and mass for the calculation of $p$; \eg Perseus has a relatively large separation but is also the most massive of all counterpart which results in $p<0.01$. Two cluster counterparts, Leo and Pisces, have $p>0.05$ (grey circles).

\section{Conclusion}
\label{sec:conclusion}
This work presents the Local Universe Model, a metric to gauge the quality of constrained simulations by comparing how similar a given simulation is, be it random or constrained, to the actual Universe. This is done by examining the proximity of candidate simulated cluster to a set of prominent rich clusters of the nearby Universe. The LUM  is applied to an ensemble of 100 DM-only cosmological simulations constrained by the Cosmicflows-3 data of peculiar velocities.

The model calculates the probability, from a random unconstrained simulation, of finding a halo at a given radial separation or closer to a random point in space (which could be the location of an observed cluster) for a range of mass cuts. This is effectively the null hypothesis which the constrained simulations are tested against.

A halo candidate from the constrained simulation is then quantitatively evaluated by calculating how likely it is that a halo of equal mass would be found at a given separation or closer in a random simulation using the one-tailed $p$-value. This $p$-value depends on the simulated halo mass and separation to the observational cluster position. The best cluster counterpart is selected to be the halo with the lowest $p$. It is assumed here that each cluster counterpart detection is independent, \ie the detection of a Virgo counterpart is not affected by the detection of any of the other cluster counterparts.

For all clusters, except Ophiuchus and Virgo, cluster counterparts with a minimum mass cut of $10^{14}$ \Msunh are found for $\gtrsim 50\%$ of the simulation for $p<0.05$ (which is 2$\sigma$ for a Gaussian distribution). This detection rate is reduced for lower $p$ thresholds. Good Ophiuchus counterparts are detected less often since its environment does not contain a density peak, as seen in the QL density, because it is not well constrained. The low detection rate of Virgo counterparts is clearly affected by the $10^{14}$ \Msunh mass cut, which is too close to the mass of the simulated Virgo counterparts. All other cluster counterparts have a wide distribution in mass except for Virgo which is contained within the low end of the mass distribution in Fig.~\ref{fig:sepmass} (right). Such low mass haloes require a very small separation to have a low $p$ that satisfies the threshold, making such counterparts unlikely to occur often. Inspection of the QL overdensity shows that the environment is well constrained which suggests that this region struggles to produce haloes above the minimum mass cut in the constrained simulations.

The LUM can also be used to evaluate the best constrained simulation by calculating the sum of $p$ for all cluster counterparts, choosing the simulation with the smallest sum, and/or counting the number of cluster counterparts below a $p$ threshold. Out of the 100 constrained simulations, the simulation with a seed of 159 was chosen. The majority of cluster counterparts in this simulation are within 10 \Mpch of the observed cluster position, where Virgo is the closest with 1.83 \Mpch. A few cluster counterparts have larger separations due to the fact that the constrained simulation has overdensities peaks, where massive haloes live, that are offset from the observed cluster positions.

The visual comparison between the density of the 159 simulation and the QL show that many of the general features match, \eg the ring-like overdensity on the SGX-SGZ plane contained within $\pm$100 \Mpch, and the arch-like feature in the SGY-SGZ plane containing Virgo, Coma and Leo. But also the underdense void regions such as between Virgo and Perseus in the SGX-SGY plane, and in the centre of the SGX-SGZ plane, are recovered. This agreement suggests that by selecting the appropriate clusters and constraining their positions, the large scale structure of the Local Universe can also be recovered.

In this work, we have presented a model that uses observational positions of clusters to evaluate the quality of constrained simulations. However, observational data has more to offer than positions. The LUM could be improved by including observation mass estimates of the clusters. Currently, the S159 simulation contains a Virgo counterpart that is more massive than the Coma counterpart. Including mass constraints would allow the relative mass distributions to match as well as their positions. This could potentially also be extended to include velocity information.

\section*{Acknowledgements}
This work has been done within the framework of the Constrained Local UniversE Simulations (CLUES) project. SP and NIL acknowledges financial support from the Deutsche Forschungs Gemeinschaft joint Polish-German research project  LI 2015/7-1 (LUSTRE). YH has been partially supported by the Israel Science Foundation grant ISF 1358/18. WAH is supported by research grants funded by the National Science Center, Poland, under agreements no. 2018/30/E/ST9/00698, 2018/31/G/ST9/03388, and 2020/39/B/ST9/03494.
\section*{Data Availability}

The simulation data used in this work are available upon reasonable request to SP.



\bibliographystyle{mnras}
\bibliography{main}




\section{Appendix A}
\label{sec:appendixa}
\begin{figure}
    \centering
    \includegraphics[width=\columnwidth]{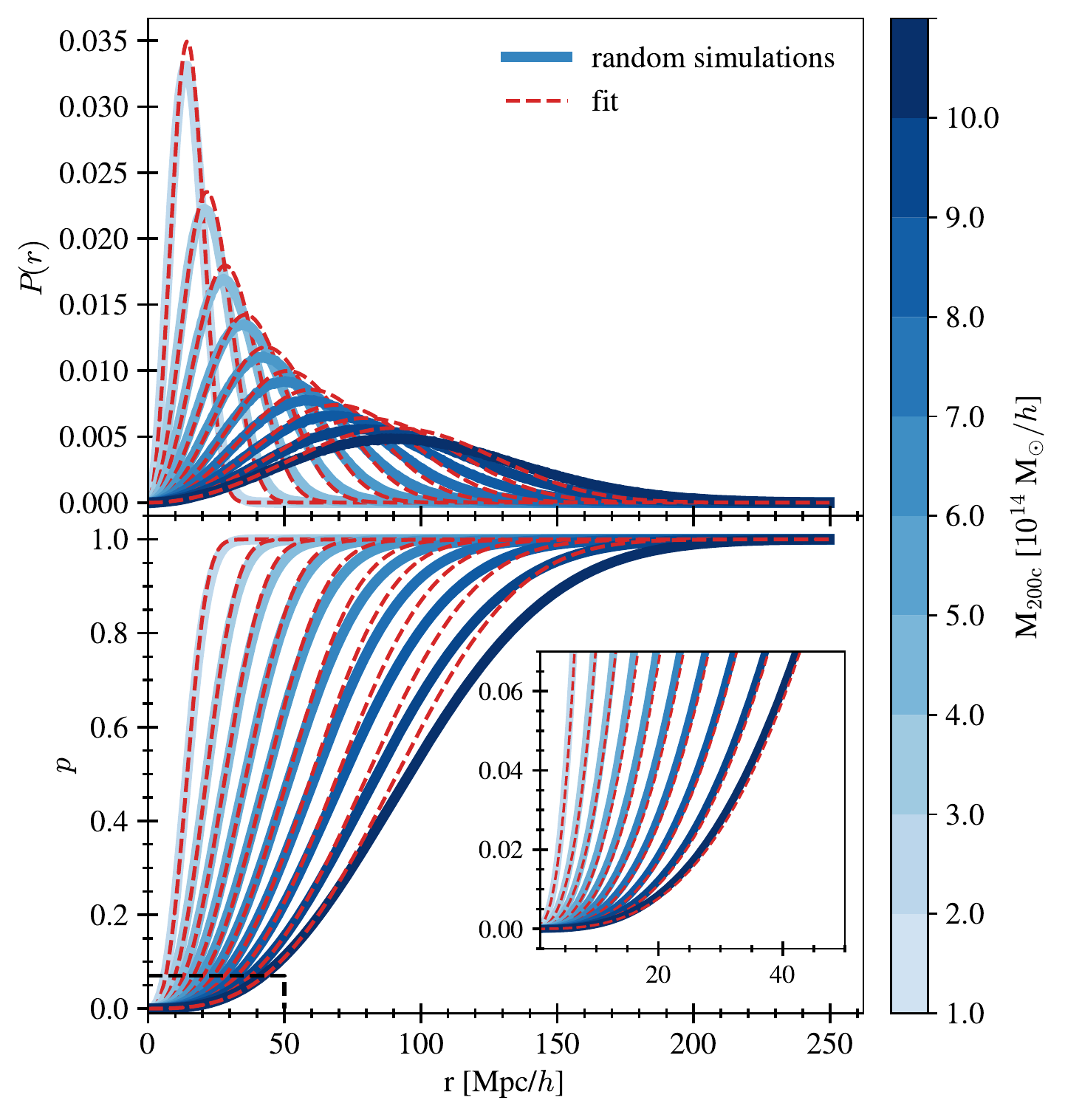}
    \caption{The PDF of the separation between the closest halo and a random point, \ie the probability of finding a halo at a given separation within a given mass cut (top). The PDFs are calculated from random, unconstrained simulations (blue) and fitted (red dashed). The $p$-value for different mass cuts (bottom) are calculate by taking the cumulative PDF, the CDF. An insert shows a zoom-in on the distributions of $p$ for small values to show the quality of the fits in the range that is important for the LUM, represented by the black dashed area.}
    \label{fig:appendixpdf}
\end{figure}

This appendix covers in more detail the fitting of the data for the LUM. As mentioned in the main text, the separation between 10000 random points and their closest halo are measure for all 100 random simulations. Fig.~\ref{fig:appendixpdf} shows the measured probability density functions (PDF) (top) for haloes with different lower mass cuts (shaded blue lines), the same as Fig.~\ref{fig:separationpdf}. Each of the 10 PDFs are fit using Equation~\ref{eq:analytic} (red dashed), prioritising data at small separations. Again, the insert shows the region of interest for the LUM where the fits are good.

\begin{figure}
    \centering
    \includegraphics[width=\columnwidth]{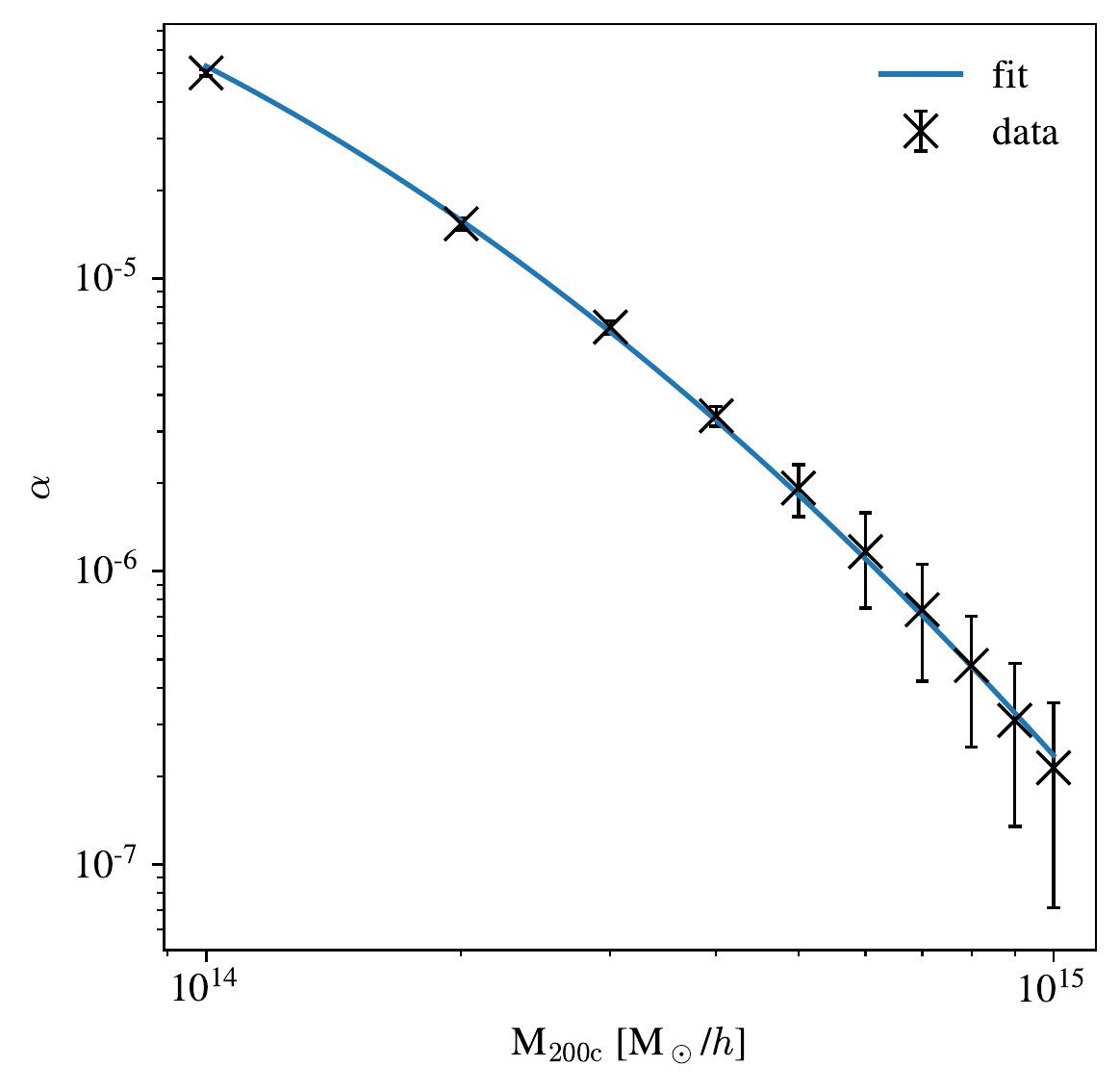}
    \caption{The data (black crosses) shows the $\alpha$ parameter fit to each PDF in Fig~\ref{fig:appendixpdf} as a function of M$_{\rm 200c}$. These are then fit with a second order polynomial (blue line) resulting in Equation~\ref{eq:alpha}. The error bars signify the standard deviation in $\alpha$ if each PDF of the 100 simulations where fit individually.}
    \label{fig:malpha}
\end{figure}

Since the functional form of the fit in Equation~\ref{eq:analytic} depends on a single parameter, $\alpha$, Fig~\ref{fig:malpha} shows the dependence of $\alpha$ on M$_{\rm 200c}$. Here, ever data point is a fit to one of the PDFs. These data are fitted using a second order polynomial (blue line) to get the function in Equation~\ref{eq:alpha}. Now the full pipeline uses two fits; a function returning an $\alpha$ from a value of  M$_{\rm 200c}$, and that $\alpha$ is used to generate a PDF. The final prediction of the PDFs from these sets of fits are shown in Fig.~\ref{fig:separationpdf} where an insignificant offset is visible from the true PDFs.

A final note about the function form of Equation~\ref{eq:analytic}. This function is the analytically derived form of the PDF distribution for a uniformly distributed set of points, for which it is exact. In this form the parameter $\alpha$ is actually the number density of the points. However, this function is not strictly appropriate since haloes are clustered. The clustering skews the PDF and creates a long tail. This is the reason why the function is a worse fit over the entire range of separations for high mass haloes because they are more clustered. The general shape of the distribution is still correct however, and is used as a fitting function instead. If the skew due to the clustering were to be accounted for, the PDFs could be generated analytically from the number density of haloes give their mass, \ie from the halo mass function.

\bsp    
\label{lastpage}
\end{document}